\documentclass[aps,prd,twocolumn,groupedaddress,amssymb,eqsecnum,showpacs,epsfig,nofootinbib]{revtex4}
\usepackage{graphicx}
\usepackage{bm}
\usepackage{dcolumn}
\usepackage{amsmath}
\usepackage{hyperref}

\usepackage{amsmath}
\usepackage{amssymb}

\numberwithin{equation}{section}
\usepackage{epsfig}
\usepackage{subfigure}

\def \lleq {\lower0.9ex\hbox{ $\buildrel < \over \sim$} ~}
\def \ggeq {\lower0.9ex\hbox{ $\buildrel > \over \sim$} ~}

\def \omm  {\Omega_{0 {\rm m}}}

\def \beq  {\begin{equation}}
\def \eeq  {\end{equation}}
\def \ber  {\begin{eqnarray}}
\def \eer  {\end{eqnarray}}
\def\bq{\begin{equation}}
\def\nq{\end{equation}}
\def\bqr{\begin{eqnarray}}
\def\nqr{\end{eqnarray}}

\begin{document}
\newcommand{\newc}{\newcommand}

\newc{\be}{\begin{equation}}
\newc{\ee}{\end{equation}}
\newc{\ba}{\begin{eqnarray}}
\newc{\ea}{\end{eqnarray}}
\newc{\bea}{\begin{eqnarray*}}
\newc{\eea}{\end{eqnarray*}}
\newc{\D}{\partial}
\newc{\daa}{\left(\frac{\Delta \alpha}{\alpha}\right)}
\newc{\ie}{{\it i.e.} }
\newc{\eg}{{\it e.g.} }
\newc{\etc}{{\it etc.} }
\newc{\etal}{{\it et al.}}
\newc{\lcdm }{$\Lambda$CDM }
\newc{\dmu}{\left(\frac{\Delta \mu (z)}{\bar \mu (z)}\right)}
\newcommand{\nn}{\nonumber}
\newc{\ra}{\rightarrow}
\newc{\lra}{\leftrightarrow}
\newc{\lsim}{\buildrel{<}\over{\sim}}
\newc{\gsim}{\buildrel{>}\over{\sim}}
\newcommand{\dslash}{D\!\!\!\!/}
\newcommand{\bx}{{\bf x}}
\newcommand{\bn}{{\bf n}}
\newcommand{\bk}{{\bf k}}
\newcommand{\dd}{{\rm d}}
\def\ga{\mathrel{\raise.3ex\hbox{$>$\kern-.75em\lower1ex\hbox{$\sim$}}}}
\def\la{\mathrel{\raise.3ex\hbox{$<$\kern-.75em\lower1ex\hbox{$\sim$}}}}

\title{Is there Correlation between Fine Structure and Dark Energy Cosmic Dipoles?}
\author{Antonio Mariano}\email{antonio.mariano@le.infn.it}
\affiliation{Department of Mathematics and Physics,
University of Salento \& INFN, Via Arnesano, 73100 Lecce, Italy\\
Physics Division, School of Technology,
Aristotle University, 54124 Thessaloniki, Greece}

\author{Leandros Perivolaropoulos}
\email{leandros@uoi.gr}
\affiliation{Department of Physics, University of Ionnina, Greece}

\date{\today}

\begin{abstract}
We present a detailed analysis (including redshift tomography) of the cosmic dipoles in the Keck+VLT quasar absorber and in the Union2 SnIa samples. We show that the fine structure constant cosmic dipole obtained through the Keck+VLT quasar absorber sample at $4.1\sigma$ level is anomalously aligned with the corresponding dark energy dipole obtained through the Union2 sample at $2\sigma$ level. The angular separation between the two dipole directions is $11.3^\circ \pm 11.8^\circ$. We use Monte Carlo simulations to find the probability of obtaining the observed dipole magnitudes with the observed alignment, in the context of an isotropic cosmological model with no correlation between dark energy and fine structure constant $\alpha$. We find that this probability is less than one part in $10^6$. We propose a simple physical model (extended topological quintessence) which naturally predicts a spherical inhomogeneous distribution for both dark energy density and fine structure constant values. The model is based on the existence of a recently formed giant global monopole with Hubble scale core which also couples non-minimally to electromagnetism. Aligned dipole anisotropies would naturally emerge for an off-centre observer for both the fine structure constant and for dark energy density. This model smoothly reduces to \lcdm for proper limits of its parameters. Two predictions of this model are (a) a correlation between the existence of strong cosmic electromagnetic fields and the value of $\alpha$ and (b) the existence of a dark flow on Hubble scales due to the repulsive gravity of the global defect core (`Great Repulser') aligned with the dark energy and $\alpha$ dipoles. The direction of the dark flow is predicted to be towards the spatial region of lower accelerating expansion. Existing data about the dark flow are consistent with this prediction.
\end{abstract}

\maketitle
\section{Introduction}
According to the cosmological principle, the Universe is homogeneous and isotropic on scales larger than a few hundred Mpc. The main source of evidence which supports this assumption comes from the Cosmic Microwave Background (CMB) which appears to be isotropic to a high degree up to a dipole term which is assumed to be due to our motion with respect to the CMB frame. However, there has been some recent observational evidence which could be interpreted as a hint for deviations from large scale statistical isotropy. Such evidence includes
alignment of low multipoles in the CMB angular power spectrum \cite{Copi:2010na}, large scale velocity flows \cite{Watkins:2008hf,Kashlinsky:2008ut} and large scale alignment in the QSO optical polarisation data \cite{Hutsemekers:2005iz} (see Ref.~\cite{Ciarcelluti:2012pc} for an interesting related theoretical model). These effects appear to persist on scales of 1 Gpc or larger and could constitute early hints for a deviation from the FLRW metric on large cosmological scales and the existence of a cosmological preferred axis. This possibility is further enhanced by the fact that the anisotropy directions implied by these observations appear to be abnormally close to each other~\cite{Antoniou:2010gw}.

The above hints for cosmological anisotropy have motivated searches for  deviations from the cosmological principle by considering the angular distribution of luminosity distances of Type Ia supernovae (SnIa) in the redshift range $z\in \left[0.015,1.4\right]$ \cite{Antoniou:2010gw,Colin:2010ds,Cooke:2009ws,Blomqvist:2008ud,Cooray:2008qn,
Gupta:2010jp,Schwarz:2007wf,Campanelli:2010zx,Cai:2011xs}. Even though all these studies are consistent with isotropy, in many of them, a mild evidence ($1\sigma-2\sigma$) of anisotropic expansion was found \cite{Antoniou:2010gw,Colin:2010ds,Cooke:2009ws,Cai:2011xs,Schwarz:2007wf} mainly coming from low redshift data, while in others \cite{Blomqvist:2008ud,Gupta:2010jp,Campanelli:2010zx} no evidence of anisotropy was found. The inability of the later studies to pick up any anisotropy is  perhaps due to the methods and data used which were not sensitive enough to particular types of anisotropy.

Additional hints for such possible deviations from the cosmological principle have recently been obtained by the angular distribution of the fine structure constant $\alpha$ in the redshift range $z\in \left[0.2223,4.1798\right]$ as measured by the quasar absorption line spectra using the many multiplet method \cite{King:2012id}. If in the case of SnIa the dipole anisotropy was mild (about $1-2\sigma$), in the case of the fine structure constant the anisotropy has been found to be significant ($4.1 \sigma$).

Some earlier studies had claimed possible variation of the fine structure `constant' with time \cite{Webb:1998cq}. This possibility has led to extensive theoretical modelling in the literature so far \cite{varconst,Chiba:2011bz} with emphasis on the possible connection of this variation with dark energy (quintessence)\cite{alpha-quint}. However, there has been comparatively less interest in the  possibility of spatial variation of $\alpha$ (see however \cite{Olive:2012ck,Olive:2010vh} for recent studies) and its connection with dark energy.

The anisotropy analysis of Ref.~\cite{Antoniou:2010gw} for the SnIa sample was based on the Union2 dataset \cite{Amanullah:2010vv} which consists of $557$ SnIa. A hemisphere comparison method was used to find the hemisphere pair with maximal anisotropy with respect to \lcdm fits. The maximum anisotropy direction was found to be towards $(l,b)= (309^\circ, 18^\circ)$ but the magnitude of this dark energy anisotropy was found to be consistent with statistical isotropy at the $2\sigma$ level. Similar results were obtained in Ref.~\cite{Colin:2010ds} where a redshift tomography also revealed that most of the contribution to the mild dark energy dipole comes from the low redshift SnIa.

The anisotropy analysis of the fine structure constant $\alpha$ \cite{King:2012id} is based on a large
sample of quasar absorption-line spectra (295 spectra) obtained using UVES (the Ultraviolet and Visual
Echelle Spectrograph) on the VLT (Very Large Telescope) in Chile and also previous observations at the Keck Observatory in Hawaii. An apparent variation of $\alpha$ across
the sky was found. It was shown to be well fit  by an angular dipole model $\daa=A \cos\theta + B$ where $\theta$ is the angle with respect to a preferred axis and $A,B$ are the dipole magnitude and an isotropic monopole term. The dipole axis was found to point in the direction
$(l,b)= (331^\circ, -14^\circ)$ and the dipole amplitude $A$ was found to be $A = (0.97\pm 0.21)\times 10^{-5}$. The statistical significance over an isotropic model was found to be at the $4.1\sigma$ level. The analysis of Ref.~\cite{King:2012id} has received criticism \cite{carroll} based mainly on the fact that its quasar sample combines two datasets (Keck and VLT) with different systematic errors  which have a small overlapping subset and cover opposite hemispheres on the sky. The axis connecting these two hemispheres has similar direction with the direction of the obtained dipole. The response of the authors of Ref.~\cite{King:2012id} was based on the fact that in the equatorial region of the dipole,
where both the Keck and VLT samples contribute a number of absorbers, there is no evidence for inconsistency between Keck and VLT.

The controversy about the possible problems in the analysis of Ref.~\cite{King:2012id} and the angular proximity between the dark energy axis of Ref.~\cite{Antoniou:2010gw} and the $\daa$ axis of Ref.~\cite{King:2012id} constitutes the motivation to analyse both the SnIa and the quasar datasets in a similar and consistent manner. Thus we re-analysed both datasets and fit them to the same dipole+monopole ansatz of the form $A \cos\theta + B$. This type of anisotropy fit is different from the corresponding SnIa fits of previous studies.
Our goal is to address the following questions:
\begin{enumerate}
\item
What are the best fit dipoles (magnitudes $A$ and directions in galactic coordinates) for the Union2 and Keck+VLT samples? What is the angle between the two dipole directions?
\item
How likely is it to obtain these dipole magnitudes in the context of an isotropic underlying model? How likely is it to obtain the observed angle between the dipoles if the two underlying models were isotropic and uncorrelated? We address these questions by producing a large number of Monte Carlo isotropic datasets simulating the Union2 and the Keck+VLT samples under the assumption of isotropic and uncorrelated underlying models. We then compare the obtained probability distributions for the dipole magnitudes and angles with the observed magnitudes and angle.
\item
How do the answers to the above questions change if we consider three different redshift slices (bins) for each dataset (low, medium and high redshift) with approximately equal number of datapoints in each bin? Is there a particular redshift range where the dark energy and the fine structure dipoles are more prominent and how is this range related with the quality of the data in each bin?
\end{enumerate}
These questions are addressed in detail in the following sections. In particular, the structure of this paper is the following: in the next section we derive the magnitudes and directions of the best fit dark energy and fine structure  dipoles for the full Union2 and Keck+VLT datasets thus addressing the above question 1. We also perform $10^4$ Monte Carlo simulations of the Union2 and Keck+VLT  datasets based on an isotropic best fit \lcdm model and on an isotropic best fit monopole model respectively. We then use these simulations to address the above question 2. In section~\ref{sec:z_tomography} we perform a redshift tomography to address question 3 and find the redshift range where the dipoles appear to be more pronounced. In section~\ref{sec:physical_mech} we  discuss a physical model that could reproduce the observed dipole alignment. Finally, in section~\ref{sec:conclusions} we conclude, summarise our basic results and discuss future prospects of the present work.

\section{Cosmic Dipoles: Data Analysis and Monte Carlo Simulations}
\subsection{Fine Structure Constant Dipole}
\label{sec:fine_structure_dipole}
The full Keck+VLT sample consists of $295$ quasar absorption line spectra in a redshift range $z\in \left[0.2223,4.1798\right]$. It has been analysed in detail in Ref.~\cite{King:2012id} where the redshift of each absorber is presented along with fine structure constant deviation $\daa=\frac{\alpha_z - \alpha_0}{\alpha_0}$ where $\alpha_z$ is the value of $\alpha$ measured at redshift $z$ using the many multiplet method \cite{Webb:1998cq} and $\alpha_0$ is the value of $\alpha$ measured in the laboratory. The positions of the quasars in equatorial coordinates are also presented.

In order to fit the Keck+VLT dataset to a dipole anisotropy we proceed as follows:
\begin{itemize}
\item
We convert the equatorial coordinates of each quasar to galactic coordinates.
\item
We find the Cartesian coordinates of the unit vectors $\hat n_i$ corresponding  to each quasar with galactic coordinates $(l,b)$. We thus have
\begin{align}
\hat n_i = \cos(b_i)  \cos(l_i) \hat i + \cos(b_i) \sin(l_i) \hat j +
\sin(b_i) \hat k  \label{hatni}
\end{align}
\item
We use the dipole+monopole angular distribution model \be \daa=A \cos\theta + B \label{dipmod} \ee where $\cos\theta$ is the angle with the dipole axis defined by the vector \be \vec D \equiv c_1 \hat i + c_2 \hat j + c_3 \hat k \label{vecd} \ee  such that \be \hat n_i \cdot \vec D = A \cos \theta_i \label{acostheta} \ee
We fit the Keck+VLT dataset to a dipole anisotropy model (\ref{dipmod}) using the maximum likelihood method i.e.\@ minimising
\be
\chi^2({\vec D},B)=\sum_{i=1}^{295}  \frac{\left[\daa_i - A \cos\theta_i  - B \right]^2}{\sigma_i^2 + \sigma_{rand}^2}
\label{chi2webb}
\ee
where $\daa_i$ and $\sigma_i$ are obtained from the Keck+VLT dataset \cite{King:2012id} and $\sigma_{rand}$ is an internal random error, assumed to be the same for all data points and representing an estimate of the aggregation of all additional  random errors. We fix the value of $\sigma_{rand}$ by requiring that at the best fit $\chi^2(\vec D,B)$ per degree of freedom is about unity. The required value of $\sigma_{rand}$ is $1.0 \times 10^{-5}$ in agreement with the corresponding value used in Ref.~\cite{King:2012id}.
\item
The magnitude and direction of the best fit dipole in galactic coordinates is obtained from the best fit $c_i$ coordinates (e.g.\@ $A=\sqrt{c_1^2 +c_2^2+c_3^2}$) and the corresponding $1\sigma$ errors are obtained using the covariance matrix approach.
\end{itemize}

\begin{figure*}[t]
\centering
\includegraphics[scale=1]{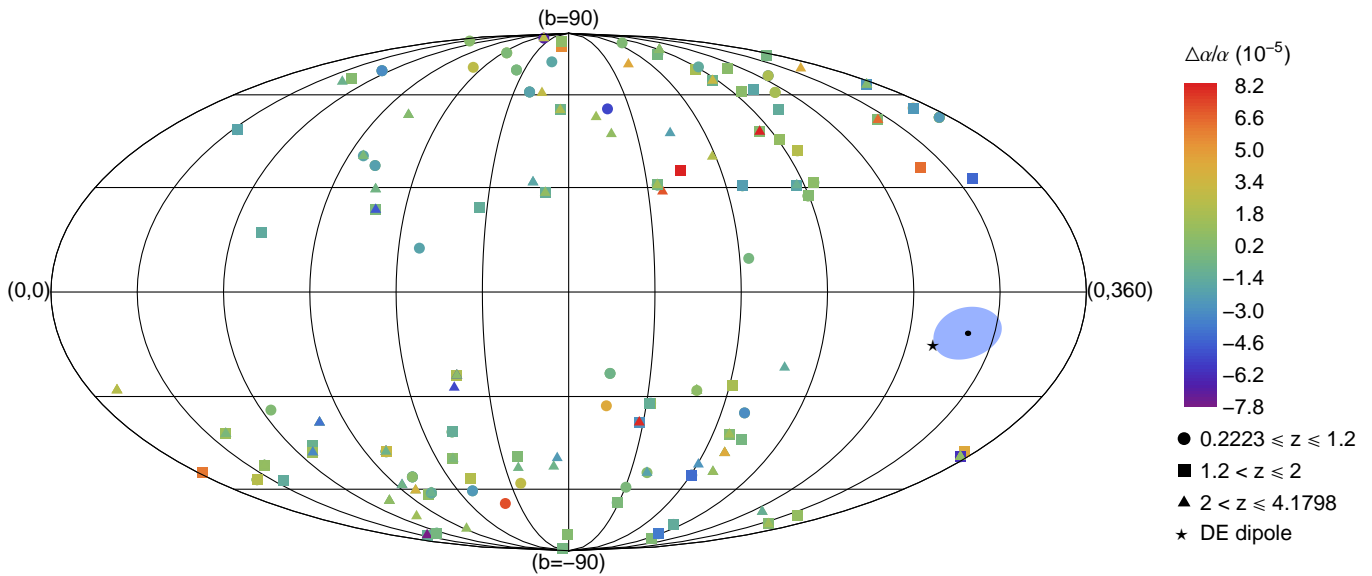}
\caption{Keck+VLT datapoints and $\alpha$-dipole direction.
Datapoints in three different redshift bins are represented with different
shapes. For comparison the direction of the Dark Energy dipole obtained from
the best fit of the Union2 data is shown with a star. The light blue blob represents the
1-$\sigma$ error on the $\alpha$-dipole direction.}
\label{fig:Webbdata}
\end{figure*}

Our result for the best fit dipole direction and magnitude is consistent with the corresponding results of Ref.~\cite{King:2012id} (see also Table~\ref{tab:WebbDipole}). We find $A_{fs}=(1.02\pm 0.25) \times 10^{-5}$ with direction ($b=-11.7^\circ \pm 7.5^\circ$, $l=320.5^\circ \pm 11.8^\circ$) while for the best fit monopole term we have $B_{fs}=(-2.2\pm 1.0)\times 10^{-6}$. This result shows that the isotropic model $A=0$ is more than $4\sigma$ away from the best fit value of the dipole magnitude. The Keck+VLT dataset along with the best fit direction of the dipole in galactic coordinates is shown in Fig.~\ref{fig:Webbdata}. By the definition of $\daa$ and the construction of the dipole model, the obtained dipole direction shown in Fig.~\ref{fig:Webbdata} is the direction towards larger values of the fine structure constant $\alpha$.

\begin{figure}[t]
\centering
\includegraphics[scale=1]{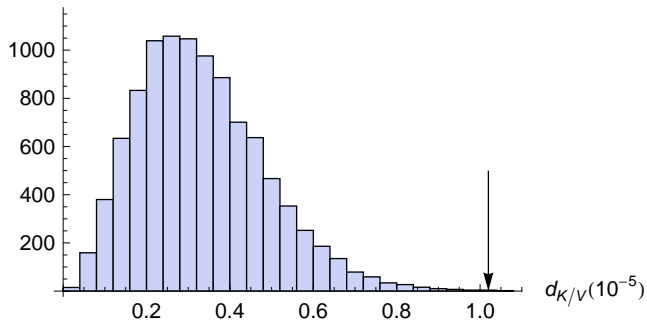}
\caption{Distribution of $\alpha$-dipole magnitudes obtained from the Monte Carlo simulation.
The arrow points to the position of the observed best fit value for the $\alpha$-dipole
magnitude.}
\label{fig:WebbMCmag}
\end{figure}

In an effort to better analyse the above results for the best fit dipole and its errors we have constructed a Monte Carlo simulation obtained from the Keck+VLT dataset under the assumption of an isotropic monopole model. Such a simulation is aimed at providing the probability distribution of the dipole magnitude and direction under the assumption of an isotropic monopole model and through that, the probability of obtaining the actually measured values in the context of an isotropic model. In order to construct the Monte Carlo simulation we proceed as follows:
\begin{itemize}
\item
We define a Gaussian random selection function $g(\mu,\sigma)$ which returns a random number from a Gaussian probability distribution with mean $\mu$ and variance $\sigma^2$.
\item
We fit the Keck+VLT dataset to an isotropic monopole model obtained from eq.~(\ref{dipmod}) by setting $A=0$. We find for the best fit monopole term: $B_{fs-m}=(-0.19 \pm 0.10)\times 10^{-5}$.
\item
We construct the isotropic Monte Carlo version of the Keck+VLT dataset by keeping fixed the direction of each quasar and assigning to each absorber an isotropic randomised fine structure constant variation obtained as:
\be
\daa_i^{MC}=g(B_{fs-m},\sigma_i)+g(0,\sigma_{rand})
\label{mcwebb}
\ee
\item
We construct $10^4$ such Monte Carlo datasets and obtain the probability distribution of the dipole magnitude as well as the corresponding dipole directions. We thus find the number of isotropic datasets that have a dipole magnitude larger than the observed value of the dipole magnitude.
\end{itemize}
The obtained probability distribution of the dipole magnitudes is shown in
Fig.~\ref{fig:WebbMCmag} along with the actually observed value of $A$.
None of the $10^4$ isotropic Monte Carlo datasets had a dipole magnitude as
large as the one observed (or larger). We thus conclude that the probability
to obtain the observed dipole magnitude of the Keck+VLT dataset in the context
of an isotropic model is less than $0.01\%$ ($3.9\sigma$) in agreement with the
covariance matrix error and with the result of Ref.~\cite{King:2012id} where the
value $4.1\sigma$ was obtained.

\subsection{Dark Energy Dipole}
We perform a similar dipole+monopole
fit using the Union2 data. Instead of $\daa$, which corresponds to fine
structure constant deviations from its earth measured value, we use the
distance modulus deviation from its best fit \lcdm value \be \dmu
\equiv \frac{\bar \mu(z)-\mu(z)}{\bar \mu(z)} \label{dmu} \ee where
$\bar \mu$ is the best fit distance modulus in the context of \lcdm.
The $557$ SnIa data points of the Union2 dataset are given in terms of
the distance
 moduli \be \mu_{obs}(z_i)\equiv m_{obs}(z_i) -
M \label{mug}\ee
 where $m_{obs}$ is the apparent magnitude of each
SnIa and $M$ is the absolute magnitude assumed to be common for all
SnIa after proper calibration.
 Assuming a \lcdm parametrisation of
the expansion rate
 \be H(z)^2 =
 H_0^2 [\omm (1+z)^3 + (1-\omm)]
\ee
 the best fit distance modulus $\bar \mu(z)$ is determined by
minimising

\be
\chi^2 (\omm,\mu_0)= \sum_{i=1}^{557}
\frac{\left[\mu_{obs}(z_i) - \mu_{th}(z_i)\right]^2}{\sigma_{\mu \;
i}^2}
\label{chi2isotr}
\ee
where $\sigma_{\mu \; i}^2$
are the distance modulus uncertainties which include both the observational and the intrinsic random magnitude scatter. The theoretical distance
modulus is defined as
\be
\mu_{th}(z_i)\equiv m_{th}(z_i) - M =5
log_{10} (D_L (z)) +\mu_0 \label{mth}
\ee
where $\mu_0$ is a constant related to the Hubble parameter $H_0\equiv 100\;h$ km/(sec $\cdot$ Mpc) \cite{Sanchez:2009ka} by
\be
\mu_0= 42.38 - 5 log_{10}h \label{mu0}
\ee
and
\be
D_L (z)= (1+z) \int_0^z dz'\frac{H_0}{H(z';\omm)} \label{dlth1}
\ee
is the Hubble free luminosity distance. A minimisation of $\chi^2 (\omm,\mu_0)$ using the Union2 dataset leads to the best fit parameter values $\omm=0.269\pm 0.020$ and $\mu_0=43.16\pm0.01$ which completely specify $\bar \mu(z_i)$ and therefore \be \left(\frac{\Delta \mu (z_i)}{\bar \mu (z_i)}\right)_{obs}\equiv \frac{\bar \mu(z_i) - \mu(z_i)}{\bar \mu(z_i)} \label{dmmudef} \ee for all Union2 datapoints.

\begin{figure*}[t]
\centering
\includegraphics[scale=1]{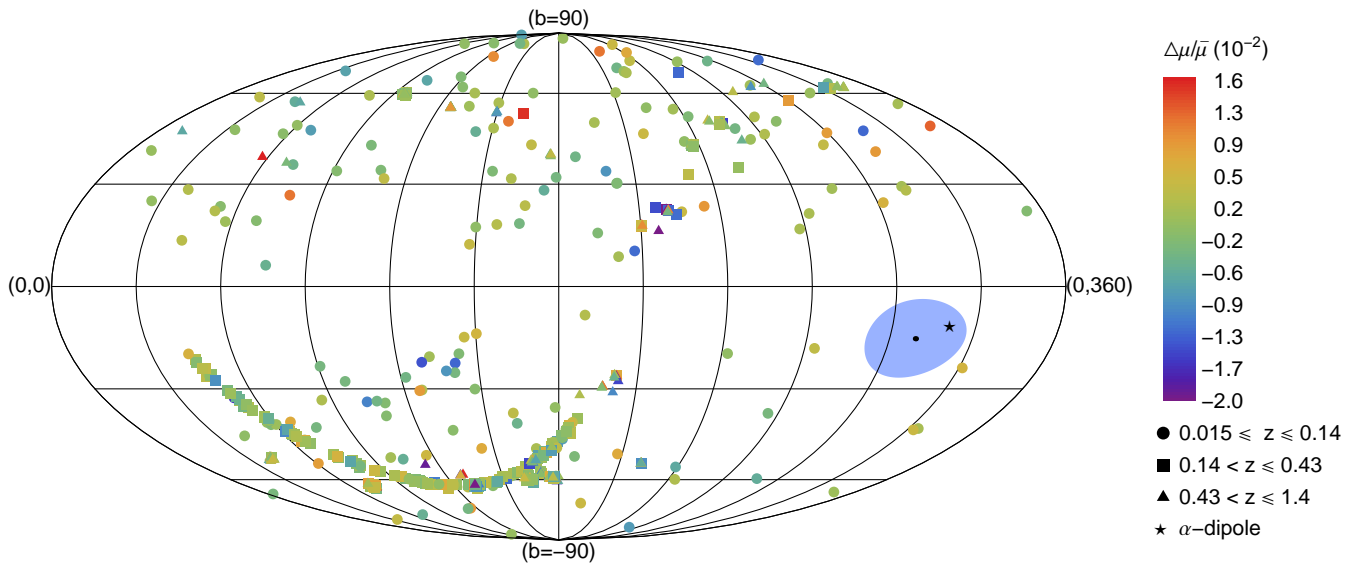}
\caption{Union2 datapoints and Dark Energy dipole direction.
Datapoints in three different redshift bins are represented with different
shapes. For comparison the direction of the $\alpha$-dipole obtained from
the best fit of the Keck-VLT data is shown with a star.
The light blue blob represents the
1-$\sigma$ error on the Dark Energy dipole direction.}
\label{fig:Union2data}
\end{figure*}

We now perform the same analysis as for the Keck+VLT data, replacing the $\daa$ datapoints by the $\dmu$ datapoints. In the SnIa we set $\sigma_{rand}=0$ since the random  intrinsic magnitude scatter has already been included in the distance moduli errors $\sigma_i$. We find the direction of the dark energy dipole to be ($b=-15.1^\circ \pm 11.5^\circ$, $l=309.4^\circ \pm 18.0^\circ$). The magnitudes of the dipole and monopole terms are found to be
\ba
A_{de}&=&(1.3\pm 0.6) \times 10^{-3} \label{dedip}\\
B_{de}&=&(2.0\pm 2.2) \times 10^{-4} \label{demon}
\ea
The statistical significance of the dark energy dipole is at the $2\sigma$ level (significantly smaller that the $4\sigma$ of the fine structure constant dipole) but its direction is only $11^\circ$ away from the corresponding direction of the fine structure constant dipole. The direction of the dark energy dipole along with the Union2 data $\left(\frac{\Delta \mu (z_i)}{\bar \mu (z_i)}\right)_{obs}$ are shown in Fig.~\ref{fig:Union2data} in galactic coordinates. The proximity of the two dipole directions is also made apparent in the same plot as well as by comparing with Fig.~\ref{fig:Webbdata}.

The direction of the dipole in Fig.~\ref{fig:Union2data} points towards brighter SnIa compared to best fit isotropic \lcdm. This implies less accelerating expansion in that direction only if $H_0$ (which is related to $\mu_0$ in eq.~(\ref{mth})) is assumed to be isotropic. This assumption was not made in Ref.~\cite{Antoniou:2010gw} where $\mu_0$ was simultaneously fit along with $\omm$ in each hemisphere. However, in the discussion of section~\ref{sec:physical_mech} we will assume isotropic $\mu_0$ and therefore lower acceleration in the direction of brighter SnIa.

\begin{figure}[t]
\centering
\includegraphics[scale=1]{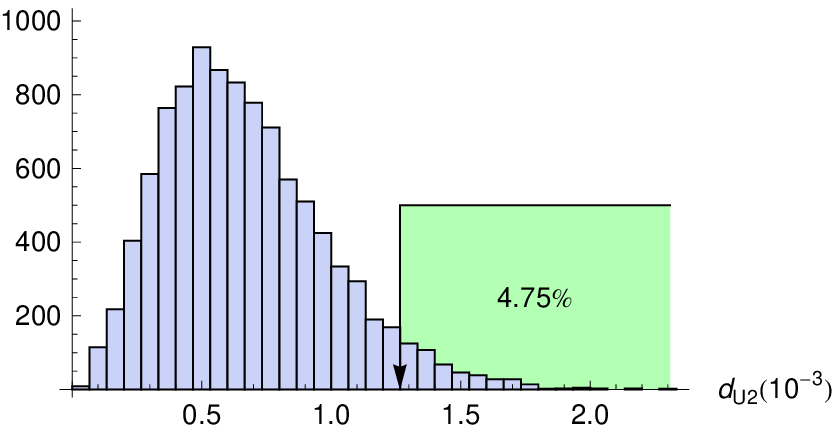}
\caption{Distribution of dark Energy dipole magnitudes obtained from the Monte Carlo simulation.
The arrow points to the position of the observed best fit value and the light green area
indicates fraction of the Monte Carlo datasets that give a dipole magnitude
bigger than the observed best fit one.}
\label{fig:Union2MCMag}
\end{figure}

\begin{figure}[t]
\centering
\includegraphics[scale=1]{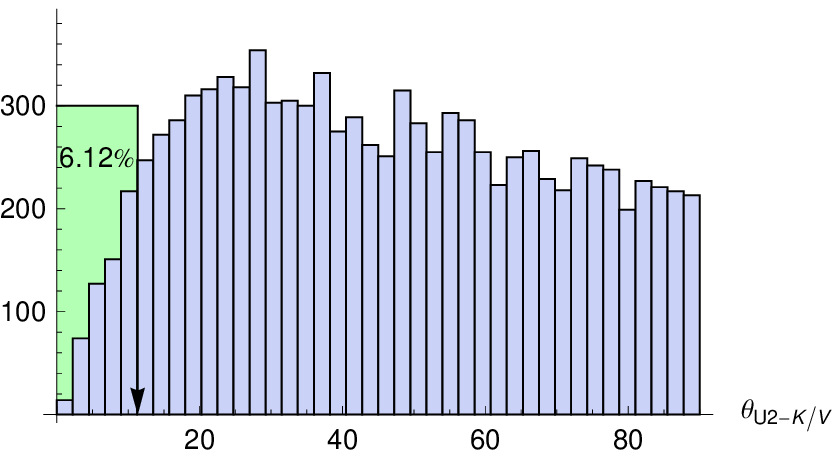}
\caption{Angular distances between the observed $\alpha$-dipole direction and the dipole
direction obtained from the Monte Carlo simulations on the Union2 data. The arrow
points to observed angular distance value and the light green area represents the Monte Carlo
datasets that give an angular distance smaller than the observed one.}
\label{fig:Union2MCangdist}
\end{figure}

\begin{figure}[t]
\centering
\includegraphics[scale=1]{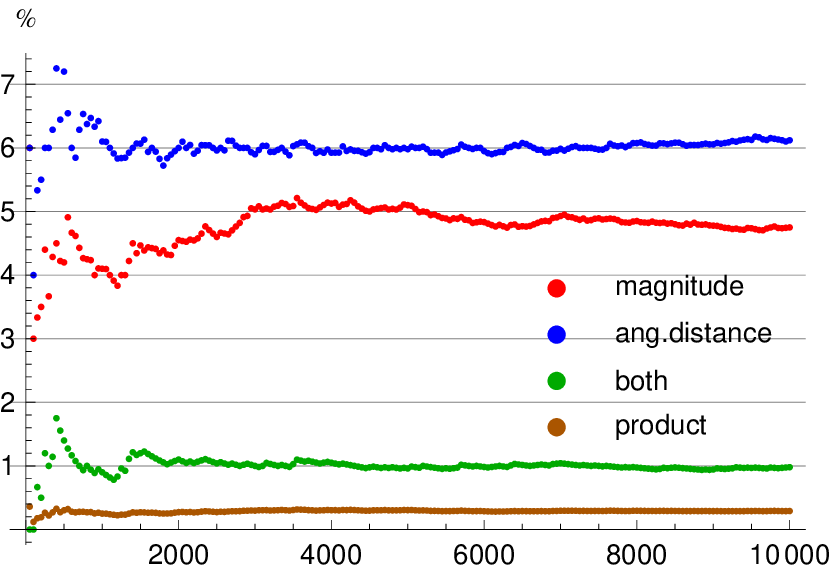}
\caption{Percentage of Union2 Monte Carlo dataset satisfying different
constraints as a function of the number of Monte Carlo datasets considered.
The points labelled as ``magnitude'' represent the fraction of Monte Carlo
datasets that give a dipole
magnitude larger than the observed one. Those labelled as ``ang.\@ distance''
represent the fraction of Union2 Monte Carlo datasets that have an angular distance from the observed $\alpha$-dipole smaller than the observed angular distance.
The label ``both'' refers to the fraction of Monte Carlo datasets that satisfy
both the previous constraints. With ``product'' we label the points that represent the product of the first two percentages.}
\label{fig:Union2MCInc}
\end{figure}

In an effort to determine the likelihood of the observed dark energy dipole magnitude combined with its angular proximity to the fine structure dipole we have performed a Monte Carlo simulation consisting of $10^4$ Union2 datasets constructed under the assumption of isotropic \lcdm. Thus the distance modulus of point $i$ is given by
\be
\mu_{MC}(z_i)=g(\bar \mu(z_i),\sigma_i)
\label{mcsnia}
\ee
where $g$ is the Gaussian random selection function defined in the previous subsection and $\bar \mu(z_i)$ is the best fit distance modulus of the Union2 data in the context of \lcdm at redshift $z_i$. It is thus straightforward to construct $\left(\frac{\Delta \mu (z_i)}{\bar \mu (z_i)}\right)_{MC}$ for each Monte Carlo dataset and obtain its best fit dipole direction and magnitude. In Fig.~\ref{fig:Union2MCMag} we show the probability distribution of the dark energy dipole magnitude in the context of isotropic \lcdm along with the observed dipole magnitude indicated by an arrow. As expected from eq.~(\ref{dedip}) only $4.75\%$ of the simulated isotropic datasets had a dark energy dipole magnitude larger than the observed value. This is consistent with eq.~(\ref{dedip}) which indicates that the statistical significance of the existence of a dark energy dipole is about $2\sigma$.
In Fig.~\ref{fig:Union2MCangdist} we show the probability distribution of the angular distance of the isotropic simulated dipoles from the observed fine structure constant dipole discussed in section~\ref{sec:fine_structure_dipole}. Only $6.12\%$ of the Monte Carlo datasets had such an angular distance smaller than the observed one. The probability for a Monte Carlo isotropic Union2 dataset to have both a dipole magnitude larger than the observed one and an angular separation from the fine structure dipole smaller than the observed one is $0.98\%$. This is larger than the anticipated value of $0.0612\times 0.0475=0.29\%$ due to the nonuniform distribution of the SnIa in the sky. The convergence of these probabilities as we increase the number of Monte Carlo simulated isotropic datasets is shown in Fig.~\ref{fig:Union2MCInc}. Clearly the number of simulated datasets considered ($10^4$) is enough to achieve the convergence of the required probabilities. We estimate the combined probability that both dipoles have magnitudes larger than the observed and angular separation smaller than the observed in the context of isotropic underlying models to be less than $0.01\% \times 0.98\% \simeq 0.0001\%$ where the first factor comes from the magnitude of the fine structure constant dipole estimated in the previous subsection.

\section{Redshift Tomography}
\label{sec:z_tomography}
In the previous section we have shown that the dipole anisotropy model provides a significantly better fit than the isotropic model for both the fine structure constant and for the dark energy spatial distributions. We also demonstrated that the two dipole directions show a remarkable coincidence. In this section we focus on identifying the redshift ranges in which these effects are more prominent. We use two approaches: a redshift bin approach and a variable upper redshift cutoff approach. In the redshift bin approach, we divide each dataset in three redshift bins of approximately equal number of datapoints and perform an analysis similar to that of the previous section in each bin. Thus we compare the results of each bin with respect to the quality of data (errorbar sizes), the dipole magnitudes and the dipole directions. In the variable upper redshift approach we start with truncated datasets with an upper redshift cutoff consisting of about 1/2 of the datapoints. Then we increase the upper redshift cutoff in five steps so that in the final step the almost full dataset is obtained. We analyse each one of the six cumulative dataset parts with respect to their dipole magnitudes and their directions.

\begin{figure*}[t]
\centering
\includegraphics[scale=1]{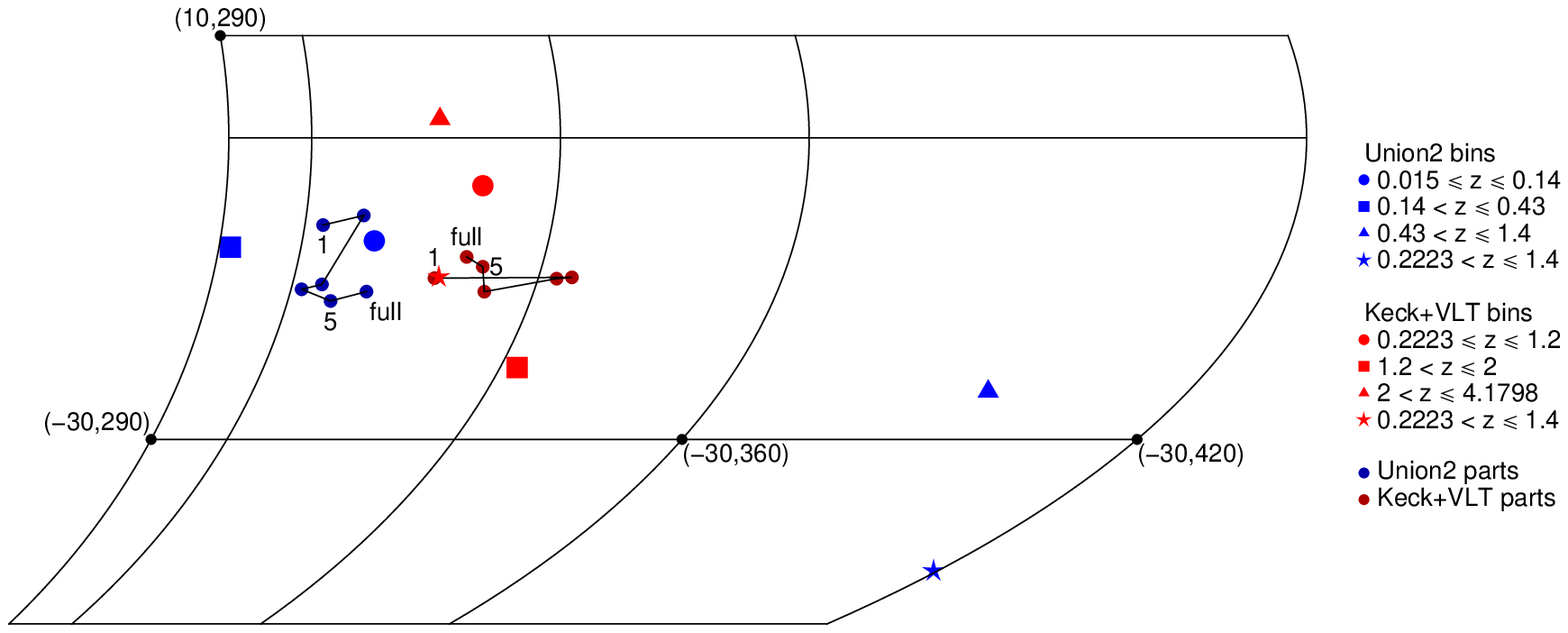}
\caption{Fine structure $\alpha$ and Dark Energy dipole directions for the different redshift
bins. The `stars' denote the bins corresponding to the redshift range that is common to the Keck+VLT and Union2 samples. For this range however, the dipole uncertainty obtained from the Union2 data is very large (see last line of Table (\ref{tab:Union2Dipole}). }
\label{fig:DipoleDirs}
\end{figure*}

In Table~\ref{tab:WebbDipole} we focus on the Keck+VLT sample and show the redshift ranges of each redshift bin and of each one of the six cumulative redshift parts. For each redshift range we show the corresponding best fit monopole magnitude, the dipole magnitude,  the direction of the best fit dipole in galactic coordinates and its angular separation from dark energy dipole (obtained from the full Union2 dataset). A similar redshift tomography for the Union2 dataset is presented in Table~\ref{tab:Union2Dipole} in which we also consider a redshift bin that is common to the two datasets (last line). The directions of the best fit dipoles for each one of the redshift ranges considered in Tables~\ref{tab:WebbDipole} and~\ref{tab:Union2Dipole} is shown in Fig.~\ref{fig:DipoleDirs}
(the cumulative redshift parts are separately connected according to increasing redshift cutoff). The uncertainties shown in Tables~\ref{tab:WebbDipole} and~\ref{tab:Union2Dipole} are obtained using the covariance matrix approach. We have checked that they are in good agreement with the corresponding $1\sigma$ errors obtained from the Monte Carlo simulations.
\begin{table*}[t]
\centering
\begin{tabular}{c|c|c|c|c|c|c}
 & $m_{K/V} (10^{-6})$ & $d_{K/V} (10^{-5})$ & $b_{d_{K/V}}(^\circ)$  & $l_{d_{K/V}}(^\circ)$ & $\theta_{K/V-U2}(^\circ)$ & datapoints \\
\hline
$0.2223 \le z \le 4.1798$ & $-2.2\pm1.0$ & 1.02 $\pm$ 0.25 & -11.7 $\pm$ 7.5 & 320.5 $\pm$ 11.8 & 11.3 $\pm$ 11.8 & 295 \\
$0.2223 < z \le 1.2$   & $-3.4\pm1.8$ & 0.8 $\pm$ 0.5 & -4.7 $\pm$ 16.8 & 320.9 $\pm$ 27.5 & 15.4 $\pm$ 25.2 & 94 \\
$1.2 < z \le 2$        & $-2.7\pm1.6$ & 0.63 $\pm$ 0.41 & -22.7 $\pm$ 20.1 & 332.2 $\pm$ 33.2 & 22.8 $\pm$ 28.5 & 103 \\
$2 < z \le 4.1798$        & $-1.5\pm2.1$ & 1.9 $\pm$ 0.4 & 1.8 $\pm$ 8.7 & 315.5 $\pm$ 11.3 & 18.0 $\pm$ 9.5 & 98\\
$0.2223 \le z \le 1.4$ & $-3.0\pm1.5$ & 0.8 $\pm$ 0.4 & -13.8 $\pm$ 14.8 & 317.3 $\pm$ 24.3 & 7.7 $\pm$ 23.7 & 125\\
$0.2223 \le z \le 1.62$& $-4.3\pm1.4$ & 0.51 $\pm$ 0.35 & -13.7 $\pm$ 20.7 & 334.1 $\pm$ 34.0 & 24.0 $\pm$ 33.4 & 152 \\
$0.2223 \le z \le 1.9$ & $-3.9\pm1.2$ & 0.7 $\pm$ 0.3 & -13.8 $\pm$ 14.3 & 332.3 $\pm$ 23.7 & 22.2 $\pm$ 23.3 &  184 \\
$0.2223 \le z \le 2.1$ & $-2.7\pm1.1$ & 0.7 $\pm$ 0.3 & -15.1 $\pm$ 11.7 & 323.9 $\pm$ 19.0 & 14.0 $\pm$ 18.5 & 208 \\
$0.2223 \le z \le 2.45$& $-2.3\pm1.1$ & 0.9 $\pm$ 0.3 & -12.6 $\pm$ 9.3 & 322.8 $\pm$ 14.9 & 13.2 $\pm$ 14.9 & 242 \\
$0.2223 \le z \le 2.7$ & $-2.4\pm1.0$ & 0.95 $\pm$ 0.27 & -12.5 $\pm$ 8.2 & 319.4 $\pm$ 13.2 & 10.0 $\pm$ 13.1 & 269 \\
\end{tabular}
\caption{Keck+VLT data: Monopole, dipole magnitude and direction and  angular distance from the
Dark Energy dipole in several redshift ranges. The
angular distance with respect
 to the Union2 dipole is referred to the full redshift
case for the Union2 dataset. We don't include the common range bin
$(0.2223 < z \le 1.4)$ since it differs from the fifth line in the table
only by one datapoint (with $z=0.2223$).}
\label{tab:WebbDipole}
\end{table*}

\begin{table*}[t]
\centering
\begin{tabular}{c|c|c|c|c|c|c}
 & $m_{U2} (10^{-4})$ & $d_{U2} (10^{-3})$ & $b_{d_{U2}}(^\circ)$ & $l_{d_{U2}}(^\circ)$ & $\theta_{U2-K/V}(^\circ)$ & datapoints \\
\hline
$0.015 \le z \le 1.4$  & $2.0\pm 2.2$ & 1.3 $\pm$ 0.6 & $-15.1$ $\pm$ 11.5 & 309.4 $\pm$ 18.0 & 11.3 $\pm$ 17.3 & 557 \\
$0.015 < z \le 0.14$   & $2.6\pm 3.4$ & 1.7 $\pm$ 0.8 & $-10.1$ $\pm$ 15.1 & 308.8 $\pm$ 22.8 & 11.6 $\pm$ 22.1 & 184 \\
$0.14 < z \le 0.43$    & $2.6\pm 5.6$ & 1.2 $\pm$ 1.9 & $-10.7$ $\pm$ 28.7 & 291.4 $\pm$ 37.2 & 28.6 $\pm$ 36.7 & 186 \\
$0.43 < z \le 1.4$     & $0.7\pm 4.3$ & 0.9 $\pm$ 0.8 & $-25.1$ $\pm$ 30.6 & 34.3 $\pm$ 75.7 & 70.6  $\pm$ 68.7 & 187 \\
$0.015 \le z \le 0.23$ & $3.3\pm 2.9$ & 1.8 $\pm$ 0.7 & $-8.5$  $\pm$ 12.4 & 302.2 $\pm$ 16.6 & 18.3 $\pm$ 16.0 & 239 \\
$0.015 \le z \le 0.31$ & $3.8\pm 2.9$ & 1.9 $\pm$ 0.7 & $-7.6$  $\pm$ 11.6 & 307.0 $\pm$ 14.7 & 13.9 $\pm$ 13.8 & 292 \\
$0.015 \le z \le 0.41$ & $3.0\pm 2.7$ & 1.8 $\pm$ 0.7 & $-14.4$ $\pm$ 10.3 & 303.6 $\pm$ 14.4 & 16.6 $\pm$ 14.1 & 352 \\
$0.015 \le z \le 0.51$ & $2.2\pm 2.6$ & 1.4 $\pm$ 0.7 & $-14.9$ $\pm$ 12.7 & 301.3 $\pm$ 18.8 & 18.9 $\pm$ 18.2 & 406 \\
$0.015 \le z \le 0.64$ & $2.1\pm 2.4$ & 1.4 $\pm$ 0.6 & $-16.0$ $\pm$ 11.0 & 305.3 $\pm$ 16.9 & 15.4 $\pm$ 16.2 & 464 \\
$0.015 \le z \le 0.89$ & $2.2\pm 2.3$ & 1.4 $\pm$ 0.6 & $-15.6$ $\pm$ 10.4 & 309.8 $\pm$ 16.0 & 11.1 $\pm$ 15.3 & 519 \\
$0.2223 < z \le 1.4 $  & $1.2\pm 2.5$ & 0.51$\pm$ 0.48& $-44.0$ $\pm$ 62.5 & 59.3  $\pm$ 147.6 & 88.2 $\pm$ 110.6 & 319 \\
\end{tabular}
\caption{Union2 data: Monopole, dipole magnitude and direction and angular distance from the
$\alpha$-dipole in several redshift ranges. The
angular distance respect
 to the $\alpha$-dipole is referred to the full
redshift case for the Keck-VLT dataset.}
\label{tab:Union2Dipole}
\end{table*}

\begin{table*}[t]
\centering
\begin{tabular}{c|c|c|c|c|c|c|c}
                       & $m_{K/V} (10^{-6})$ & $d_{K/V} (10^{-5})$ & $b_{d_{K/V}}(^\circ)$ & $l_{d_{K/V}}(^\circ)$ & $\textrm{RA}_{d_{K/V}}(hr)$ & $\textrm{dec}_{d_{K/V}}(^\circ)$ & dp  \\
\hline
$0.2223 \le z \le 1.6$ & -3.9$\pm$1.1        & 0.57 $\pm$ 0.26     & -16.4 $\pm$ 15.1      & 336.4 $\pm$ 22.9      & 18.1 $\pm$ 2.0              & -57.3 $\pm$ 20.9                 & 148 \\
$1.6 < z \le 4.1798$   & 1.1$\pm$1.4         & 1.39 $\pm$ 0.35     & -10.5 $\pm$ 7.8       & 325.5 $\pm$ 11.9      & 16.6 $\pm$ 1.4              & -63.0 $\pm$ 10.2                 & 145 \\
\end{tabular}
\caption{
Monopole, dipole magnitude and direction in the low and high redshift ranges for
the Keck+VLT data.
The results have been obtained fitting the data using the same values of $\sigma_{rand}$
as in~\cite{King:2012id} (three values) and removing the two outliers
as identified by~\cite{King:2012id}. These results are almost identical with those of Ref.~\cite{King:2012id} which provides a good test of our analysis. We have checked that using a single value $\sigma_{rand}=1.0$ for all datapoints (as done in the rest of our analysis) leads to consistent results and affects mainly the error bars which become somewhat larger.
}
\label{tab:WebbDipoleKingComp}
\end{table*}


The following comments can be made based on the results shown in Tables~\ref{tab:WebbDipole} and~\ref{tab:Union2Dipole}:
\begin{itemize}
\item
The redshift bin with the smallest $1\sigma$ errors (best data quality) for the Union2 data is the low redshift bin ($z\in \left[0.015,0.14\right]$). The corresponding best data quality redshift bin for the Keck+VLT dataset is the high redshift bin ($z\in \left[2,4.1798\right]$).
\item
These best quality redshift bins also have the best dipole alignment and the most statistically significant deviation of the best fit dipole magnitudes from isotropy. It is therefore important to improve the quality of data in the other redshift bins in order to clarify whether the dipole trend is also strong in these bins where the data quality is lower.
\end{itemize}

\begin{figure}[t]
\centering
\includegraphics[scale=1]{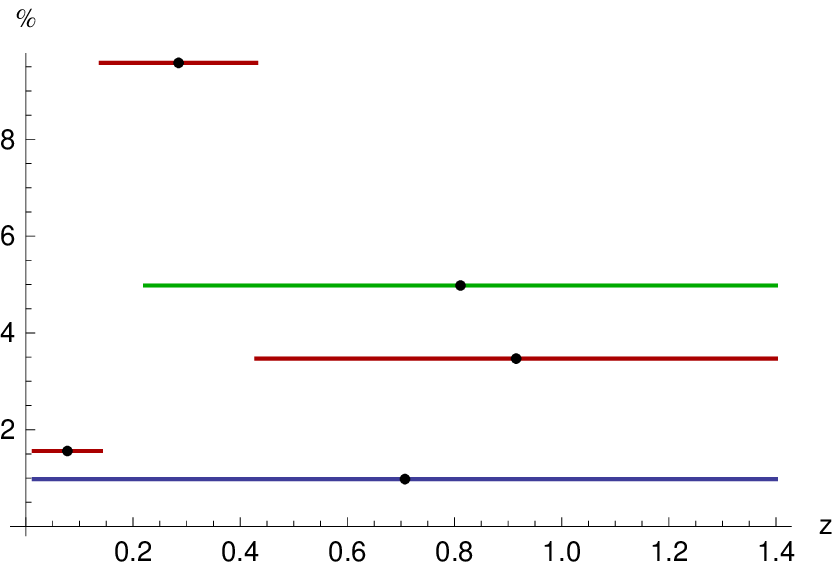}
\caption{Percentage of the Monte Carlo Union2 datasets that give both a Dark
Energy dipole larger than the observed value and an angular distance between
the Dark Energy dipole and the $\alpha$-dipole smaller than the observed angular
distance. The result is plotted for the full and partial redshift bins.}
\label{fig:BothProbability}
\end{figure}

\begin{figure}[t]
\centering
\includegraphics[scale=1]{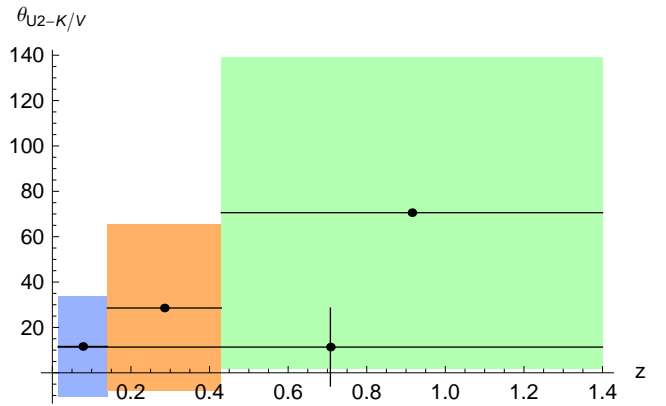}
\caption{Angular distances (with errors) between the Dark Energy dipole obtained
in the different redshift bins and the $\alpha$-dipole obtained from the full
redshift range Keck+VLT data. Notice that the lowest Union2 redshift bin dipole has the best alignment with the Keck+VLT dipole and also has the smallest error.}
\label{fig:AngDistU2}
\end{figure}

\begin{figure}[t]
\centering
\includegraphics[scale=1]{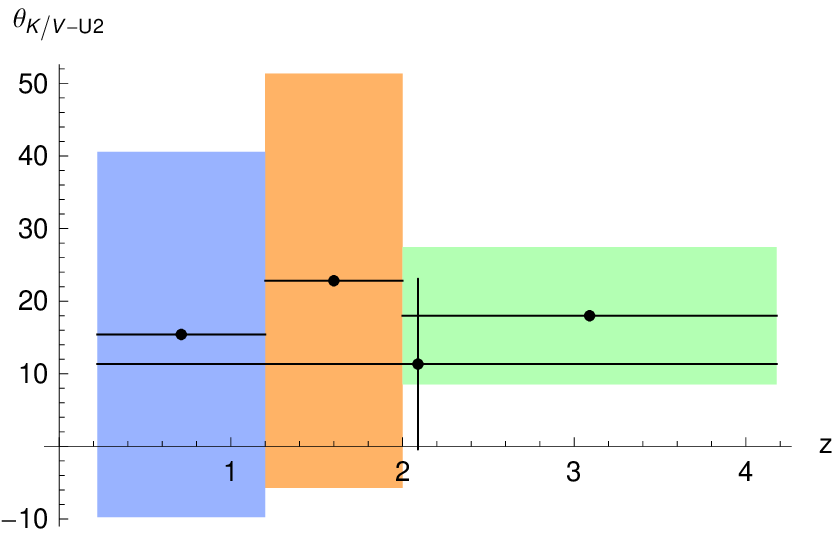}
\caption{Angular distances (with errors) between the $\alpha$-dipole obtained
in the different redshift bins and the  Dark Energy dipole obtained from the full
redshift range Union2 data. Notice that the alignment of all Keck+VLT redshift bin dipoles with the full Union2 dipole is consistent with each other and similar to the alignment of the full Keck+VLT dataset.}
\label{fig:AngDistWebb}
\end{figure}

The above points are also demonstrated in Figs.~\ref{fig:BothProbability}, \ref{fig:AngDistU2} and \ref{fig:AngDistWebb}. For each one of the redshift bins considered, we show in Fig.~\ref{fig:BothProbability} the fraction of isotropic Union2 datasets that exceed the observed dipole dark energy magnitude and also have a smaller angular distance from the Keck+VLT dipole than the actually observed angular distance. Clearly, this fraction is significantly lower for the lowest redshift bin which implies that the dipole behaviour and alignment is most significant for this redshift bin. In Fig.~\ref{fig:AngDistU2} we show the angular separation of each Union2 redshift bin from the best fit dipole direction of the full Keck+VLT dataset, as a function of redshift range for each Union2 bin. Clearly, the lowest redshift bin which also has the smallest angular separation error is the one that has its dipole best aligned with the Keck+VLT dipole. Finally, in Fig.~\ref{fig:AngDistWebb} we show the angular separation of each Keck+VLT redshift bin from the best fit dipole direction of the full Union2 dataset, as a function of redshift range for each bin. Clearly, the highest redshift bin has the lowest error and good alignment with the Union2 dipole. In this case however, the best fit dipole direction appears to be more consistent among the three redshift bins while the errorbars for the dipole direction are significantly smaller than the Union2 case.

The above choice of redshift bin ranges has been made by demanding approximately equal number of datapoints in each one of the three redshift bin. In Ref~\cite{King:2012id} two redshift bins were considered: a high redshift bin with $z>1.6$ and a low redshift bin with $z<1.6$. The motivation for this redshift division comes from the fact that high redshift absorber spectra are dominated by different absorption lines compared to low redshift absorption spectra. Thus a test for possible systematics could be to divide the whole sample and compare the two resulting dipoles. If the two dipoles are consistent with each other then this is an indication that no systematic errors are hidden in the different absorption lines. No systematic errors were found in Ref~\cite{King:2012id} using this approach.

In order to test the consistency of our results with those of Ref.~\cite{King:2012id} we have used the same two bins and constructed Table~\ref{tab:WebbDipoleKingComp} which is to be compared with the corresponding Table 3 of Ref.~\cite{King:2012id}. In an effort to reproduce the results of ~\cite{King:2012id} we have used the same values of $\sigma_{rand}$ and ignored two outlier datapoints. Our results are almost identical with those of Ref.~\cite{King:2012id} and this provides a good test of the validity of our analysis.

\section{Physical Mechanism: Extended Topological Quintessence}
\label{sec:physical_mech}
If the observed coincident large dipole anisotropies are due to a physical mechanism and not to systematic or statistical fluctuations, then it is of particular interest to investigate what could be a physical model that would give rise simultaneously to these coincident dipoles. Such a mechanism could involve for example an inhomogeneous scalar field which couples to electromagnetism through a non-minimal coupling and whose potential energy could provide the dark energy required for accelerating expansion. Due to negative pressure such a scalar field would tend to quickly become homogeneous and isotropic on Hubble scales. However, nontrivial topology would naturally generate sustainable inhomogeneity~\cite{BuenoSanchez:2011wr} of such a scalar field.

For a proper potential, the scale of the inhomogeneity would be the observationally required Hubble scale. In such a Hubble scale topological defect, an off centre observer would observe aligned dipoles in both dark energy and the fine structure constant. For a large enough core scale, such a defect would become effectively homogeneous and indistinguishable from \lcdm. The dipole nature of observations of off-centre observers located in spherically symmetric inhomogeneities has been discussed in detail in Refs \cite{Grande:2011hm,Alnes:2006pf}.

In the case of no coupling to electromagnetism, this mechanism was studied in detail in Ref.~\cite{BuenoSanchez:2011wr} (topological quintessence). Topological quintessence is an extension of the well known corresponding inflationary model: topological inflation \cite{topinfl}. In what follows we present some qualitative features of the {\it extended} topological quintessence model and we postpone a more detailed study for a later publication.

\begin{figure*}[t]
\centering
\subfigure[]{
\includegraphics[scale=1]{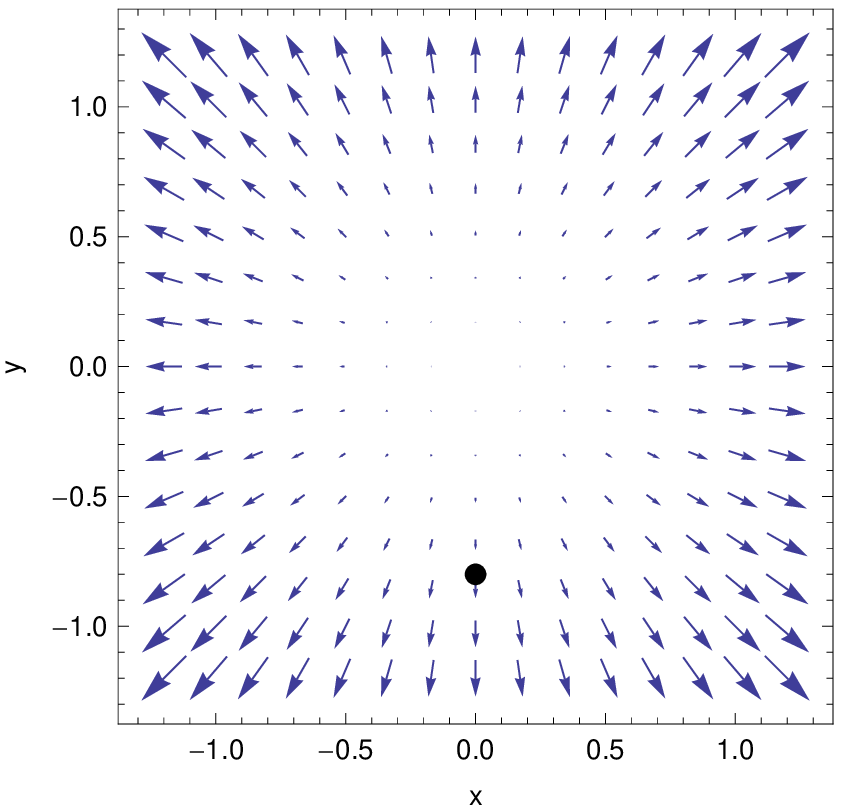}
\label{fig:MonObsA}
}
\hspace{.2cm}
\subfigure[]{
\raisebox{1cm}{\includegraphics[scale=1]{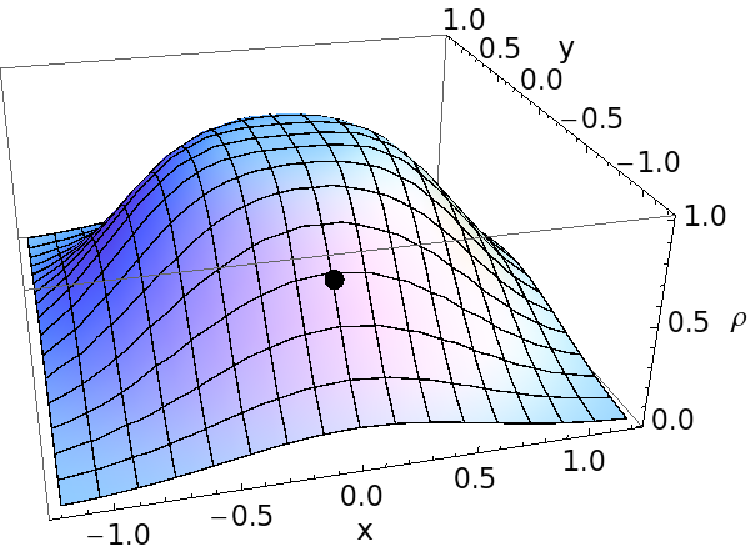}}
\label{fig:MonObsB}
}
\caption{(a) The observer location (thick dot) along with the field magnitude and direction denoted by the arrows at each point of the $x-y$ plane. Clearly, the field magnitude is smaller towards the centre of the monopole and this justifies the variation of $\alpha$ in that direction. (b)  The energy density ($\rho$) distribution of the global monopole and the location of the observer. Clearly, there is higher dark energy density towards the monopole centre and this justifies the higher acceleration rate in that direction.}
\label{fig:MonObs}
\end{figure*}

Consider the action
\begin{eqnarray}
 S =\int \left[ \frac{1}{2}M_p^2 R -\frac12(\partial_{\mu}\Phi^a)^2
 -V(\Phi) + \right. \nonumber \\
\left. + \frac{1}{4}B(\Phi) F_{\mu\nu}^2 + {\cal L}_m
 \right]\sqrt{-g}\dd^4 x,
\end{eqnarray}
where $M_p^{-2}=8\pi G$ is the reduced Planck mass, $ {\cal L}_m$ is the Lagrangian density of matter fields,  $\Phi^a ~(a=1,2,3)$ is an $O(3)$ symmetric scalar field, $B(\Phi)$ is a non-minimal coupling to electromagnetism and
\begin{equation}\label{gmptl}
V(\Phi)= {1\over 4}\lambda(\Phi^2-\eta^2)^2, ~~
\Phi\equiv\sqrt{\Phi^a\Phi^a}\,.
\end{equation}
We assume the existence of a Hubble scale global monopole formed during a recent cosmological phase transition. The vacuum energy density in the monopole core and the size of the core are determined by the two parameters of the model
$\eta$ (the vacuum expectation value) and $\lambda$ (the coupling constant).  The global monopole field configuration is described by the hedgehog ansatz
\begin{equation}\label{glmonanz}
\Phi^a=\Phi(r,t)\hat r^a\equiv
\Phi(r,t)(\sin\theta\cos\varphi,\sin\theta\sin\varphi,\cos\theta)
\end{equation}
shown in Fig. \ref{fig:MonObsA}, with boundary conditions
\begin{equation}
\Phi(0,t)=0\quad,\quad\Phi(\infty,t) =\eta\,,
\end{equation}
where $\eta$ is the scale of symmetry breaking assumed to be such that \cite{BuenoSanchez:2011wr}
\be
{{\lambda \eta^2}\over 3 H_0^2} \gsim 1
\label{dyncond}
\ee
In eq.~(\ref{glmonanz}) we have allowed for a time dependence having in mind a cosmological setup of an expanding background.
For a slowly evolving global monopole configuration, the size of the core is approximated by
\be \delta \simeq \lambda^{-1/2} \eta^{-1}\,, \label{coresc} \ee while the vacuum energy density in this core region is
\be \rho^{core} \simeq \frac{\lambda \eta^4}{4}\,. \label{rhomoncore} \ee For a core size much larger than the Hubble scale, the model reduces to \lcdm. Therefore the constraints imposed on inhomogeneous matter models\cite{Zumalacarregui:2012pq} are not applicable to this class of inhomogeneous dark energy models.

The general spherically symmetric spacetime around a global monopole may be described by a metric of the form
\begin{equation}\label{ltb}
ds^2=-dt^2+A^2(r,t)dr^2+B^2(r,t)r^2(d\theta^2+\sin^2\theta d\varphi^2).
\end{equation}
A detailed analysis of the cosmological evolution of the above metric in the presence of the global monopole and matter is presented in Ref.~\cite{BuenoSanchez:2011wr} with $B(\Phi)=1$ (see also Ref.~\cite{Grande:2011hm,topinfl}).

Consider now a non-minimal coupling of the form
\be
B(\Phi)=1-\xi \frac{\Phi^2}{\eta^2}
\label{nmcoupl}
\ee
where $\xi$ is constant. The fine structure `constant' is related to the coupling $B(\Phi)$ as
\be
\alpha(\Phi)=\frac{e_0^2}{4\pi B(\Phi)^2}
\label{alphabphi}
\ee
where $e_0$ is the bare charge that remains constant throughout the cosmological evolution. Therefore for small values of $\Phi/\eta$ we have
\be
\daa \simeq 2\xi \frac{(\Phi^2-\Phi_0^2)}{\eta^2}
\label{daaphi1}
\ee
where $\Phi_0$ is the field magnitude at the location of the observer.
The dipole directions shown in Figs.~\ref{fig:Webbdata} and \ref{fig:Union2data} correspond to higher value of $\alpha$ and lower accelerating expansion (brighter SnIa compared to \lcdm) respectively. Thus, in the extended topological quintessence picture, for an off-centre observer, this would be the direction pointing away from the global monopole core where the potential energy of the monopole is lower and the field magnitude $\Phi$ is larger. In order to have a higher value of the fine structure constant in the same direction we need $\xi>0$.

In Fig.~\ref{fig:MonObs} we illustrate the location of an off-centre observer with respect to the monopole core. In Fig.~\ref{fig:MonObsA} we plot the observer location along with the field magnitude and direction denoted by the arrows at each point of the $x-y$ plane. Clearly, the field magnitude is smaller towards the centre of the monopole and this justifies the variation of $\alpha$ in that direction. Similarly, in Fig. \ref{fig:MonObsB} we show the energy density distribution of the global monopole and the location of the observer. Clearly, there is higher dark energy density towards the monopole centre and this justifies the higher acceleration rate in that direction.

We postpone a detailed reconstruction of the global monopole potential $V(\Phi)$ and coupling $B(\Phi)$ for a later publication. A comparison of the quality of fit for different topological defect geometries could also be made (global vortex or thick domain wall). We also stress that our extended topological quintessence approach is distinct from the model of Refs.~\cite{Olive:2012ck,Olive:2010vh} where thin domain walls were considered in an effort to explain the spatial variation of $\alpha$. The approach of Refs.~\cite{Olive:2012ck,Olive:2010vh} does not address the dark energy dipole and predicts a non-dipole anisotropy for $\alpha$. The dipole anisotropy however has been shown \cite{Berengut:2012ep} to provide a better fit to the Keck+VLT data that the single wall model \cite{Olive:2010vh}. The double wall model \cite{Olive:2012ck} however, involving three additional parameters, has been shown to provide a better fit than the dipole model.

The extended topological quintessence monopole discussed above is also distinct from the {\it varying $\alpha$ defects} \cite{varadef} based on Bekensteins's theory \cite{bektheor}. According to this model, the electric charge $e$ (and therefore also the fine structure constant $\alpha$) is promoted to a dimensionless scalar field $\varphi\sim \ln e$ with zero potential and a kinetic term multiplied by a large dimensionful parameter $\omega$. This is similar to the corresponding extension of General Relativity along the lines of the Brans-Dicke theory where Newton's constant is promoted to a scalar field. As in the Brans-Dicke theory, the parameter $\omega$ is used to partly {\it freeze} the dynamics of $\varphi$ so that the charge variation in spacetime becomes consistent with observational and experimental constraints \cite{Chiba:2011bz,expcona}. The dynamics of the charge field $\varphi$ affects the dynamics of the gauge field $A_\mu$ which in turn affects the dynamics of any scalar field $\Phi$ that couples to $A_\mu$ via a gauge symmetry. Varying $\alpha$ defects \cite{varadef} are gauged defects formed if the vacuum manifold of $\Phi$ has a non-trivial homotopy group and their dynamics is {\it indirectly} affected by the dynamics of the fine structure constant (and of $\varphi$). A potential source of severe constraints for this class of defects is that they predict massive photons (spontaneous breaking of electromagnetism) in regions away from the defect core.

In contrast to these varying $\alpha$ defects, in extended topological quintessence, the defect is global and is formed by the same field that represents $\alpha$. These are global defects non-minimally coupled to electromagnetism. As a simple example in Minkowski spacetime, consider a global vortex non-minimally coupled to electromagnetism. The dynamics of the complex scalar field $\Phi$, is determined by the Lagrangian density
\be
{\cal L}\, = \left(\partial_\mu\,\Phi\right)^{*}\left(\partial^\mu\,\Phi\right)
-\frac{1}{4}\,B(\Phi)\,F_{\mu\nu}\,F^{\mu\nu} -V(\Phi) \, , \label{langvor}
\ee
The field equations obtained by variation of $\Phi^*$ and $A_\mu$ are
\be\label{eqphi}
\partial_\mu\,\partial^\mu\,\Phi\,=\,-\frac{\partial\, V}{\partial\,\Phi^\star}\,-\frac{1}{4}\,
\frac{\partial B (\Phi)}{\partial\, \Phi^*}\, \,F_{\mu\nu}\,F^{\mu\nu}\,.
\ee
and
\be\label{eqaa}
\partial_\nu\left[B(\Phi)\,F^{\mu\nu}\right]\,=0,
\ee
For a non-minimally coupled global vortex to form we set
\be
\label{mexhat}
V(\Phi) = \frac{\lambda}4 \left(\Phi^* \Phi - \eta^2 \right)^2,
\ee
and
\be
B(\Phi)=1-\xi \frac{\vert \Phi \vert ^2}{\eta^2}
\label{nmcouplvort}
\ee
We now use the global vortex the ansatz allowing for a coaxial magnetic field
\ba
\Phi &=& f(r)\,e^{i n \theta}\, , \label{fan}\\
A_\theta  &=& a(r)\, , \label{aan}
\ea
where $f(r)$ and $a(r)$ are real functions of $r$ and
all other components of $a_\mu$ are set to zero. We thus obtain the static field equations for $f(r)$ and $a(r)$ as
\begin{widetext}
\ba
&&\frac{1}{r}\frac{d}{dr}\left(r\frac{d f}{dr}\right)-
\left(\frac{n^2}{r^2}
-\frac{\eta^2\,\lambda}{2}+\frac{\lambda}{2} f^2\right)f-\frac{1}{2}
\frac{dB(f^2)}{df}
\left(\frac{1}{r}\frac{d}{dr} \left(ra\right)\right)^2=0, \label{feq}\\
&&\frac{d}{dr}\left(B(f^2)\frac{1}{r}\frac{d}{dr} \left(r a\right)\right)=0,\label{aeqmot}
\ea
\end{widetext}
since
\be
F_{\mu\nu}\,F^{\mu\nu} = 2 F^{r\theta} \, F_{r\theta} = 2 \left[ \frac1{r} \frac{d}{dr}(ra) \right]^2.
\label{error}
\ee
The corresponding energy density of the vortex is:
\be
\rho =  \left( \frac{d f}{dr}  \right)^2 + \frac{1}{2\,r^2} B(f^2) \left( \frac{d\,  (r\, a)}{dr}  \right)^2 +  \frac{n^2}{r^2} f^2  + \frac{\lambda}4
\left( f^2 - \eta^2 \right)^2.
\ee
If there is no external source of electromagnetic field and if $B(f^2)>0$ everywhere, we obtain the usual global vortex solution $f(r)=f_0(r)$, $a(r)=0$. However, if there are regions of space where $B(f^2)<0$, an instability develops which proceeds with spontaneous creation of electromagnetic field in the region where $B(f^2)<0$. For example for $B(f^2)=1-q V(f^2)/\eta^4$, an instability develops in the core, for large enough values of $q$.

If there is an external source of electromagnetic fields (e.g.\@ a localised magnetic field in the $z$ direction), then the profile of $f(r)$ will be affected in accordance with eq.~(\ref{feq}) and a local additional variation of $\alpha$ will occur. Thus, a robust prediction of this class of models is a correlation between regions of strong electromagnetic fields and variation of $\alpha$. The non-observation of such variation could impose strong constraints on the form of the coupling $B(f^2)$. The detailed investigation of these constraints and their consistency with the form of $B(f^2)$ required to explain the observed $\alpha$ dipole represents an interesting extension of this project.

\section{Conclusions}
\label{sec:conclusions}
We have used the Keck+VLT dataset and the Union2 dataset to show that the value of the fine structure constant and the rate of accelerating expansion are better described by coinciding dipoles than by isotropic cosmological models. The key feature of our analysis is that it applies identical method (fit to dipole+monopole anisotropy) to both the Keck+VLT dataset and the Union2 dataset. This consistency, combined with the apparent dipole nature of the anisotropy, has allowed a consistent comparison of the two dipoles.

Using Monte Carlo simulations and covariance matrix error estimates, we find that the probability that these coinciding dipoles are both produced in the context of a cosmological model where fine structure constant and dark energy are isotropic and uncorrelated is less than one part in $10^6$. A redshift tomography analysis dividing the two datasets in three redshift bins revealed that the highest data quality redshift bins correspond to low redshifts for the Union2 sample and high redshift for the Keck+VLT data. The dipole direction for the Keck+VLT data depends weakly on redshift while the Union2 dipole direction depends more strongly on redshift and it is the low redshift (and lowest error) bin that is best aligned with the Keck+VLT dipole. The directional uncertainty is significantly larger for the medium and higher redshift Union2 dipoles. It is therefore important to improve the quality of intermediate and high redshift SnIa data in order to further test the alignment of the dark energy dipole with the fine structure constant dipole.

An important issue that we have not addressed in the present paper is the effect of systematic errors of the Keck+VLT sample. This issue has been addressed in detail in Ref.~\cite{King:2012id} where no significant source of systematic errors was identified. The main concern has been the possibility of careless merging of the two datasets (VLT and Keck) which in principle have different systematics and effectively cover opposite hemispheres of the sky which coincide with the direction of the identified Keck+VLT dipole. The concern therefore is that the large identified dipole magnitude originates from a hidden difference in systematic errors between the VLT and Keck samples\cite{carroll}. According to Ref.~\cite{King:2012id} this does not appear to be the case for the following reasons:
\begin{itemize}
\item
the dipole directions at high and low redshifts are in agreement (this is confirmed in our study too as shown in  Figs.~\ref{fig:DipoleDirs} and \ref{fig:AngDistWebb});
\item
the directions of the dipoles fitted by the VLT and by the Keck samples separately are in agreement;
\item
the absorbers that are common to both the VLT and the Keck sample provide consistent values for $\alpha$.
\end{itemize}
Even though the above arguments of Ref.~\cite{King:2012id} are reasonable, a truly convincing analysis would involve observation of the same objects with a different telescope. This has already been done by the Subaru telescope in August 2004 \cite{Chiba:2011bz}. An analysis of these observations could provide a particularly useful independent verification of the fine structure constant dipole.

Finally we have proposed a theoretical model that has the potential to predict strong aligned dipoles for the fine structure constant and for dark energy. The model is based on a non-minimal coupling of a topologically non-trivial scalar field to electromagnetism (extended topological quintessence). In such a model, an off-centre observer with respect to the Hubble scale core of a global monopole would naturally observe large aligned dipoles for the fine structure constant and dark energy. In fact it should be possible to reconstruct both the scalar field potential and the non-minimal coupling form using the Keck+VLT and the Union2 samples.

A robust prediction of the non-minimally coupled defect model is the weak dependence of the value of $\alpha$ on the existence of local strong magnetic fields as discussed at the end of the previous section. Another interesting prediction is the existence of peculiar velocities in the direction away from the center of the global monopole due to the repulsive effects of antigravity (negative pressure) in the defect center. An off-center observer would experience this Hubble scale flow as a dipole dark flow. Such dipole dark flow has indeed been observed \cite{Kashlinsky:2008ut,Watkins:2008hf} and it is attributed to the existence of a Great Attractor which could be present on $Gpc$ scales (perhaps even at a neighboring universe \cite{MersiniHoughton:2008rq}). In our model such a dark flow could be due to a 'Great Repulser' whose role would be played by the core of the Hubble scale non-minimally coupled defect. The predicted direction of such a flow should be away from the defect core (Great Repulser) in the direction of maximum deceleration ($b=-15.1^\circ \pm 11.5^\circ$, $l=309.4^\circ \pm 18.0^\circ$) (see Table \ref{tab:Union2Dipole}). The direction of the observed dark flow is ($b=8^\circ \pm 6^\circ$, $l=287^\circ \pm 9^\circ$)\cite{Watkins:2008hf} which is consistent within $1\sigma$ with the direction of the dark energy and $\alpha$ dipoles. It also points towards the region of lower acceleration as predicted by our model. A robust prediction of our model with respect to the dark flow is that it should reverse direction at large enough redshifts as we start seeing on the 'other side' of the 'Great Repulser' (defect core).

A detailed investigation of the consistency of the above predictions with cosmological observations is an interesting extension of the present analysis.

{\bf Numerical Analysis Files:} The data, Mathematica and C++program files used for the numerical analysis files may be downloaded from \url{http://leandros.physics.uoi.gr/defsdipoles}.

\section*{Acknowledgements}
We thank Michael Murphy for clarifications regarding the Keck+VLT data of Ref.~\cite{King:2012id}. We also thank Tanmay Vachaspati for useful comments. AM was supported by the European Union under the Marie Curie Initial
Training Network ``UNILHC'' PITN-GA-2009-237920. This research has been co-financed by the European Union (European Social Fund - ESF) and Greek national funds through the Operational Program "Education and Lifelong Learning" of the National Strategic Reference Framework (NSRF) - Research Funding Program: THALIS.\@ Investing in the society of knowledge through the European Social Fund.


\begin{thebibliography}{99}

\bibitem{Copi:2010na}
  C.~J.~Copi, D.~Huterer, D.~J.~Schwarz and G.~D.~Starkman,
  arXiv:1004.5602 [astro-ph.CO].

\bibitem{Watkins:2008hf}
  R.~Watkins, H.~A.~Feldman and M.~J.~Hudson,
  arXiv:0809.4041 [astro-ph];   H.~A.~Feldman, R.~Watkins and M.~J.~Hudson,
  Mon.\ Not.\ Roy.\ Astron.\ Soc.\  {\bf 407}, 2328 (2010)
  [arXiv:0911.5516 [astro-ph.CO]].

\bibitem{Kashlinsky:2008ut}
  A.~Kashlinsky, F.~Atrio-Barandela, D.~Kocevski and H.~Ebeling,
  Astrophys.\ J.\  {\bf 686}, L49 (2009)
  [arXiv:0809.3734 [astro-ph]]; A.~Kashlinsky, F.~Atrio-Barandela and H.~Ebeling,
  arXiv:1202.0717 [astro-ph.CO];   A.~Kashlinsky, F.~Atrio-Barandela, H.~Ebeling, A.~Edge and D.~Kocevski,
  Astrophys.\ J.\  {\bf 712}, L81 (2010)
  [arXiv:0910.4958 [astro-ph.CO]].

\bibitem{Hutsemekers:2005iz}
  D.~Hutsemekers, R.~Cabanac, H.~Lamy and D.~Sluse,
  Astron.\ Astrophys.\  {\bf 441}, 915 (2005)
  [arXiv:astro-ph/0507274];
  D.~Hutsemekers, A.~Payez, R.~Cabanac, H.~Lamy, D.~Sluse, B.~Borguet and J.~R.~Cudell,
  arXiv:0809.3088 [astro-ph];
  D.~Hutsemekers and H.~Lamy,
  arXiv:astro-ph/0012182.

\bibitem{Ciarcelluti:2012pc}
  P.~Ciarcelluti,
  arXiv:1201.6096 [astro-ph.CO].

\bibitem{Antoniou:2010gw}
  I.~Antoniou and L.~Perivolaropoulos,
  JCAP {\bf 1012}, 012 (2010)
  [arXiv:1007.4347 [astro-ph.CO]].

\bibitem{Colin:2010ds}
  J.~Colin, R.~Mohayaee, S.~Sarkar and A.~Shafieloo,
  Mon.\ Not.\ Roy.\ Astron.\ Soc.\  {\bf 414}, 264 (2011)
  [arXiv:1011.6292 [astro-ph.CO]].

\bibitem{Cooke:2009ws}
  R.~Cooke and D.~Lynden-Bell,
  Mon.\ Not.\ Roy.\ Astron.\ Soc.\  {\bf 401}, 1409 (2010)
  [arXiv:0909.3861 [astro-ph.CO]].

\bibitem{Blomqvist:2008ud}
  M.~Blomqvist, E.~Mortsell and S.~Nobili,
  JCAP {\bf 0806}, 027 (2008)
  [arXiv:0806.0496 [astro-ph]]; M.~Blomqvist, J.~Enander and E.~Mortsell,
  JCAP {\bf 1010}, 018 (2010)
  [arXiv:1006.4638 [astro-ph.CO]].

\bibitem{Cooray:2008qn}
  A.~R.~Cooray, D.~E.~Holz and R.~Caldwell,
  JCAP {\bf 1011}, 015 (2010)
  [arXiv:0812.0376 [astro-ph]].

\bibitem{Gupta:2010jp}
  S.~Gupta and T.~D.~Saini,
  arXiv:1005.2868 [astro-ph.CO].

\bibitem{Schwarz:2007wf}
  D.~J.~Schwarz and B.~Weinhorst,
  Astron.\ Astrophys.\  {\bf 474}, 717 (2007)
  [arXiv:0706.0165 [astro-ph]].

\bibitem{Campanelli:2010zx}
  L.~Campanelli, P.~Cea, G.~L.~Fogli and A.~Marrone,
  Phys.\ Rev.\ D {\bf 83}, 103503 (2011)
  [arXiv:1012.5596 [astro-ph.CO]].

\bibitem{Cai:2011xs}
  R.~-G.~Cai and Z.~-L.~Tuo,
  JCAP {\bf 1202}, 004 (2012)
  [arXiv:1109.0941 [astro-ph.CO]].


\bibitem{King:2012id}
  J.~K.~Webb, J.~A.~King, M.~T.~Murphy, V.~V.~Flambaum, R.~F.~Carswell and M.~B.~Bainbridge,
  Phys.\ Rev.\ Lett.\  {\bf 107}, 191101 (2011)
  [arXiv:1008.3907 [astro-ph.CO]];
  J.~A.~King, J.~K.~Webb, M.~T.~Murphy, V.~V.~Flambaum, R.~F.~Carswell, M.~B.~Bainbridge, M.~R.~Wilczynska and F.~E.~Koch,
  arXiv:1202.4758 [astro-ph.CO].

\bibitem{Webb:1998cq}
  J.~K.~Webb, V.~V.~Flambaum, C.~W.~Churchill, M.~J.~Drinkwater and J.~D.~Barrow,
  Phys.\ Rev.\ Lett.\  {\bf 82}, 884 (1999)
  [astro-ph/9803165];   M.~T.~Murphy, J.~K.~Webb, V.~V.~Flambaum, V.~A.~Dzuba, C.~W.~Churchill, J.~X.~Prochaska, J.~D.~Barrow and A.~M.~Wolfe,
  Mon.\ Not.\ Roy.\ Astron.\ Soc.\  {\bf 327}, 1208 (2001)
  [astro-ph/0012419];   M.~T.~Murphy, J.~K.~Webb and V.~V.~Flambaum,
  Mon.\ Not.\ Roy.\ Astron.\ Soc.\  {\bf 345}, 609 (2003)
  [astro-ph/0306483].

\bibitem{varconst}
  J.~-P.~Uzan,
  Rev.\ Mod.\ Phys.\  {\bf 75}, 403 (2003)
  [hep-ph/0205340];   Y.~Fujii and S.~Mizuno,
  Int.\ J.\ Mod.\ Phys.\ D {\bf 14}, 677 (2005)
  [astro-ph/0404222];   H.~B.~Sandvik, J.~D.~Barrow and J.~Magueijo,
  Phys.\ Rev.\ Lett.\  {\bf 88}, 031302 (2002)
  [astro-ph/0107512].

\bibitem{Chiba:2011bz}
  T.~Chiba,
  Prog.\ Theor.\ Phys.\  {\bf 126}, 993 (2011)
  [arXiv:1111.0092 [gr-qc]].

\bibitem{alpha-quint}
G.~R.~Dvali and M.~Zaldarriaga,
  Phys.\ Rev.\ Lett.\  {\bf 88}, 091303 (2002)
  [hep-ph/0108217];   C.~Wetterich,
  Phys.\ Lett.\ B {\bf 561}, 10 (2003)
  [hep-ph/0301261];   L.~Anchordoqui and H.~Goldberg,
  Phys.\ Rev.\ D {\bf 68}, 083513 (2003)
  [hep-ph/0306084];  E.~J.~Copeland, N.~J.~Nunes and M.~Pospelov,
  Phys.\ Rev.\ D {\bf 69}, 023501 (2004)
  [hep-ph/0307299];   D.~-S.~Lee, W.~Lee and K.~-W.~Ng,
  Int.\ J.\ Mod.\ Phys.\ D {\bf 14}, 335 (2005)
  [astro-ph/0309316];   N.~J.~Nunes and J.~E.~Lidsey,
  Phys.\ Rev.\ D {\bf 69}, 123511 (2004)
  [astro-ph/0310882];   V.~Marra and F.~Rosati,
  JCAP {\bf 0505}, 011 (2005)
  [astro-ph/0501515];   M.~Byrne and C.~Kolda,
  hep-ph/0402075.

\bibitem{Olive:2012ck}
  K.~A.~Olive, M.~Peloso and A.~J.~Peterson,
  arXiv:1204.4391 [astro-ph.CO].

\bibitem{Olive:2010vh}
  K.~A.~Olive, M.~Peloso and J.~-P.~Uzan,
  Phys.\ Rev.\ D {\bf 83}, 043509 (2011)
  [arXiv:1011.1504 [astro-ph.CO]].

\bibitem{carroll}
  E.~Cameron and T.~Pettitt,
  arXiv:1207.6223 [astro-ph.CO];
N.~Kanekar, G.~I.~Langston, J.~T.~Stocke, C.~L.~Carilli and K.~L.~Menten,
  arXiv:1201.3372 [astro-ph.CO];
  S.~A.~Levshakov, F.~Combes, F.~Boone, I.~I.~Agafonova, D.~Reimers and M.~G.~Kozlov,
  arXiv:1203.3649 [astro-ph.CO].

\bibitem{Amanullah:2010vv}
  R.~Amanullah {\it et al.},
  Astrophys.\ J.\  {\bf 716}, 712 (2010)
  [arXiv:1004.1711 [astro-ph.CO]].

\bibitem{Sanchez:2009ka}
  J.~C.~B.~Sanchez, S.~Nesseris and L.~Perivolaropoulos,
  JCAP {\bf 0911}, 029 (2009)
  [arXiv:0908.2636 [astro-ph.CO]].

\bibitem{BuenoSanchez:2011wr}
  J.~C.~Bueno Sanchez and L.~Perivolaropoulos,
  Phys.\ Rev.\ D {\bf 84}, 123516 (2011)
  [arXiv:1110.2587 [astro-ph.CO]].

\bibitem{Grande:2011hm}
  J.~Grande and L.~Perivolaropoulos,
  Phys.\ Rev.\ D {\bf 84}, 023514 (2011)
  [arXiv:1103.4143 [astro-ph.CO]].

\bibitem{Alnes:2006pf}
  H.~Alnes and M.~Amarzguioui,
  Phys.\ Rev.\ D {\bf 74}, 103520 (2006)
  [astro-ph/0607334].

\bibitem{topinfl}
A.~Vilenkin,
  Phys.\ Rev.\ Lett.\  {\bf 72}, 3137 (1994)
  [hep-th/9402085];   A.~D.~Linde,
  Phys.\ Lett.\ B {\bf 327}, 208 (1994)
  [astro-ph/9402031];   N.~Sakai, H.~-A.~Shinkai, T.~Tachizawa and K.~-i.~Maeda,
  Phys.\ Rev.\ D {\bf 53}, 655 (1996)
  [Erratum-ibid.\ D {\bf 54}, 2981 (1996)]
  [gr-qc/9506068]; I.~Cho and A.~Vilenkin,
  Phys.\ Rev.\ D {\bf 56}, 7621 (1997)
  [gr-qc/9708005]; A.~A.~de Laix, M.~Trodden and T.~Vachaspati,
  Phys.\ Rev.\ D {\bf 57}, 7186 (1998)
  [gr-qc/9801016].

\bibitem{Zumalacarregui:2012pq}
  M.~Zumalacarregui, J.~Garcia-Bellido and P.~Ruiz-Lapuente,
  arXiv:1201.2790 [astro-ph.CO].


\bibitem{Berengut:2012ep}
  J.~C.~Berengut, E.~M.~Kava and V.~V.~Flambaum,
  arXiv:1203.5891 [astro-ph.CO].

\bibitem{varadef}
J. Magueijo, H. Sandvik and T. W. B. Kibble, Phys. Rev. D{\bf 64}, 023521 (2001); J.~Menezes, P.~P.~Avelino and C.~Santos,
  Int.\ J.\ Mod.\ Phys.\ A {\bf 21}, 3295 (2006)
  [gr-qc/0601007];   J.~Menezes, P.~P.~Avelino and C.~Santos,
  Phys.\ Rev.\ D {\bf 72}, 103504 (2005)
  [hep-ph/0509326];  J.~Menezes da Silva,
  arXiv:0808.3274 [hep-ph];


\bibitem{bektheor}
J. D. Bekenstein,  Phys. Rev. D {\bf 25}, 1527 (1982); H. B. Sandvik, J. D. Barrow, and J. Magueijo,  Phys. Rev.
Lett. {\bf 88}, 031302 (2002); D. F. Mota and J. D. Barrow,  Phys. Lett. B {\bf 81}, 141
(2004); D. F. Mota and J. D. Barrow,  Mon.
Not. Roy. Astron. Soc. {\bf 349}, 291 (2004); D. Kimberly and J. Magueijo, Phys. Lett. B {\bf 584}, 8
(2004); J. D. Barrow, J. Magueijo and H. B. Sandvik,  Phys. Rev. D{\bf 66},
043515 (2002); P. P. Avelino, C. J. A. P. Martins, and J. C. R. E. Oliveira, Phys. Rev. D{\bf 70}, 083506 (2004);   J.~D.~Barrow and S.~Z.~W.~Lip,
  Phys.\ Rev.\ D {\bf 85}, 023514 (2012)
  [arXiv:1110.3120 [gr-qc]].

\bibitem{expcona}
R. Srianand, H. Chand, P. Petitjean, and B. Aracil, Phys. Rev. Lett. {\bf 92}, 121302 (2004); K. A. Olive et al.,
Phys. Rev. D {\bf 69}, 027701 (2004); D. J. Shaw and J. D. Barrow,  Phys.
Lett. B {\bf 639}, 596 (2006); J. D. Barrow,  Phys. Rev. D {\bf 71},
083520 (2005); T. Damour, "Varying constants", gr-qc/0306023 (2003); 
\bibitem{Damour:1996zw}
  T.~Damour and F.~Dyson,
  Nucl.\ Phys.\ B {\bf 480}, 37 (1996)
  [hep-ph/9606486]; T. Damour and F. Dyson, Nucl. Phys. B {\bf 480}, 37 (1996); S. Blatt et al., Phys. Rev. Lett. {\bf 100}, 140801 (2008); J.~D.~Barrow and D.~J.~Shaw,
  Phys.\ Rev.\ D {\bf 78}, 067304 (2008)
  [arXiv:0806.4317 [hep-ph]];   J.~Magueijo, J.~D.~Barrow and H.~B.~Sandvik,
  Phys.\ Lett.\ B {\bf 549}, 284 (2002)
  [astro-ph/0202374].

\bibitem{MersiniHoughton:2008rq}
  L.~Mersini-Houghton and R.~Holman,
  JCAP {\bf 0902}, 006 (2009)
  [arXiv:0810.5388 [hep-th]].

\end{thebibliography}
\end{document}